\documentclass{ws-mpla}
\usepackage[super]{cite}
\usepackage{graphicx}
\usepackage{booktabs}
\begin{document}

\markboth{J. M. Hoff da Silva and R. Cavalcanti}{}

\catchline{}{}{}{}{}

\title{REVEALING HOW DIFFERENT SPINORS CAN BE:\\ THE LOUNESTO SPINOR CLASSIFICATION}

\author{\footnotesize J. M. Hoff da Silva}

\address{Departamento de F\'isica e Qu\'imica, Universidade Estadual Paulista-UNESP\\
Av. Ariberto Pereira da Cunha, 333, Guaratinguet\'a-SP, Brazil\\
hoff@feg.unesp.br}

\author{R. T. Cavalcanti}

\address{Departamento de F\'isica e Qu\'imica, Universidade Estadual Paulista-UNESP\\
Av. Ariberto Pereira da Cunha, 333, Guaratinguet\'a-SP, Brazil\\
rogerio.txc@feg.unesp.br}

\maketitle

\pub{Received (Day Month Year)}{Revised (Day Month Year)}

\begin{abstract}
This paper aims to give a coordinate based introduction to the so-called Lounesto spinorial classification scheme. We introduce the main ideas and aspects of this spinorial categorization in an argumentative basis, after what we delve into a commented account on recent results obtained from (and within) this branch of research. 

\end{abstract}

\ccode{PACS Nos.: 02.10.De, 02.90.+p, 03.65.Fd}

\vspace{.5cm}
\begin{quote}
{\it ... because spinors cannot be constructed by tensorial methods.}

Pertti Lounesto.
\end{quote}
\vspace{.5cm}

\section{Introduction}	

It is difficult to give the right account on the importance that spinorial fields have in the description of high energy process. Perhaps to assert that spinors describe the basic blocks of all the matter content of the universe may give a starting trigger on such an importance. 

In the scope widely used in fundamental physics, symmetry may be categorized into two great groups: on the one hand we have the so-called dynamical laws, relating interactions felt by particles, and, on the other hand, the kinematical laws, dealing with symmetries of the spacetime over which the physics takes part \cite{wig}. Spinors are important in both cases. Firstly, the right appreciation of the spacetime symmetries shows that half integer eigenvalues of the Pauli-Lubanski quadratic form, a Casimir invariant of the Poincar\`e group, are to be taken into account. These representations are indeed described by spinors. Moreover, even though spinors are not usually related to describe interaction, they do feel interactions. Even further, it is imperative to say, theories bringing the concept of spinors in its foundations, as quantum electrodynamics for instance, describe nature with an almost incredible accuracy. 

Despite of this broad applicability and usefulness in the understanding of nature, when talking about spinors one is almost naturally lead to the concept of Dirac spinors. This fact is understandable and actually quite deserved, since the very use of spinors in high energy physics is largely due to the impressive work of Dirac \cite{Dirac}. Going a little further, but perhaps in a smaller scale, Weyl and Majorana spinors are also contained in the tool box of the modern physicist. Part of the relevance of the issue to be treated here is to give a right account on different type of spinors --- including possibilities different from the mentioned --- and some of its properties as well.     

When studying spinors and the first fundamental related aspects, a crucial point appears as relevant: a single given spinor cannot be detected. Indeed, the very Lorentz group composition law related to spinors do not recover the original state when a $2\pi$ rotation are taken into account \cite{wig2,wei}. In other words a given spinor cannot be alone a physical observable. Nevertheless, composite (even) quantities of spinors can be associated to observables, thus the importance of the so-called bilinear covariants. The canonical procedure when dealing with spinor fields can be recast in the following protocol: find the dual for a given spinor field and, then, proceed writing down all the bilinear covariants associated to them. These bilinears, according to a well defined context, are physical observables quantities. This is what the widely widespread wisdom about spinor fields dictates, and it is present in almost every text book in quantum field theory. In the next section we shall only briefly report ourselves to these aspects.

A less known, but rather important, result about spinors and bilinears is that under very reasonable considerations not only the bilinears are obtained from spinors but the opposite is also true: spinor fields can be recovered from the bilinears they gave rise. This astonishing achievement is known as the inversion theorem and it is due to Takahashi \cite{taka}. We shall take advantage of this theorem, in a more suitable disguise, to envisage a spinor field classification performed by Lounesto \cite{lou}. This classification recast spinors according to the behavior of its bilinear covariants and, hence, is a physical appealing categorization of spinors. The link between different bilinears rearrangements and possible different spinors is given by the Fierz-Pauli-Kofink (FPK) identities \cite{fpk}. The FPK identities are quadratic relations to be obeyed by the bilinear covariants of a given spinor field representing, thus, important constraints to the different spinor possibilities. This paper aims to recover the aforementioned spinor categorization, calling attention to the essential aspects of the formalism without entering the tough ground of excessive mathematical technicality, but arguing in a precise manner. 

The present paper is structured as follows: in the next section, after a brief introduction to the canon about spinors culminating with the bilinear covariants, we recall an argument, due to Crawford, on the use of the inversion theorem when constrained by the FPK identities related to a certain class of spinors \cite{craw}. After that we move forward showing the classification scheme and evincing its main consequences. Section \ref{shortcuts} is devoted to give an account on the main results based on, or extending, the Lounesto spinor classification. In the final section we conclude.   

\section{The Lounesto classification scheme}\label{LCS}

There are many ways to define a spinor \cite{rol}. For the purposes of this paper, we shall consider a spinor $\psi$ as an element of $P_{SL(2,\mathbb{C})}\times_\rho \mathbb{C}^4$. Let us explain in detail this definition. Obviously, $SL(2,\mathbb{C})$ is the double covering of $SO(1,3)$ group, the group comprising Lorentz transformation. $P_{SL(2,\mathbb{C})}$ denotes the set of all inertial frames connected by $SL(2,\mathbb{C})$ transformations (an appropriate place for physics happen, indeed), i. e. connected by the invariance group. The spinor is, then, an element of such a bundle, but a rather special one, an element appearing in a four complex entries disposal ($\mathbb{C}^4$) carrying a spin 1/2 linear representation of the invariance group. This last aspect is encoded in the $\rho$ specification of the definition: $\rho$ stands for the representation, or Weyl, space comprised by right-hand (according to a generic boost) two-entries spinors $(1/2,0)$, left-hand two-entries spinors $(0,1/2)$ or a complete four spinor of $(1/2,0)\oplus(0,1/2)$. This definition may readily accommodate usual text-books spinors. In this section we shall particularize our analysis to spinors obeying the Dirac equation. 

Consider $\psi$ annihilated by the Dirac operator. Since, by definition, $\psi$ belongs to a linear representation of the Lorentz group, it must exist $S(\Lambda)$ such that $\psi'(x')=S(\Lambda)\psi(x)$, where $\Lambda$ is a Lorentz transformation matrix. From this observation it is straightforward to see that the covariance of the Dirac equation requires 
\begin{equation}
S^{-1}\gamma^{\mu}S=\Lambda^{\mu}_{\;\;\nu}\gamma^\nu.\label{1} 
\end{equation} Along this text $\gamma^\mu$ are the usual gamma matrices in the Weyl representation\footnote{The reader can easily check that the constitutive relation of the Clifford algebra $\{\gamma^\mu,\gamma^\nu\}=2\eta^{\mu\nu}$ is invariant with respect to Eq. (\ref{1}).}. 

Before evincing the full set of bilinear covariants we shall take advantage of the spinorial dual theory \cite{dva}. As we are interested in usual Dirac spinors, at first, we start defining the dual as $\psi_{Dual}=\psi^\dagger \eta$, where $\eta$ is a matrix whose elements are to be determined\footnote{Notice that there is at least one more important element of freedom. We are chosen {\it ab initio} $\psi_{Dual}=(\mathbb{I}\psi)^\dagger \eta$. Had we chosen any other operator different from the identity, we would arrive at a different dual \cite{dva}.}. Our guidance is given by the exigence that the product $(\psi_{Dual})\psi$ be a Lorentz scalar. Thus, let $p^\mu$ be the four momentum obtained from $k^\mu=(m,\lim_{|{\bf p}|\downarrow 0} \frac{{\bf p}}{|{\bf p}|})$ by a general boost. Being the entire representation space boost generator, ${\bf \kappa}$, given by a $(-i{\bf \sigma}/2,+i{\bf \sigma}/2)$  block diagonal matrix, where ${\bf \sigma}$ are the usual Pauli matrices, and the rotation generator, ${\bf \xi}$, represented by a $({\bf \sigma}/2,{\bf \sigma}/2)$ block diagonal matrix, it is fairly simple to see that under a general boost, the norm invariance $(\psi_{Dual}(p^\mu))\psi(p^\mu)=(\psi_{Dual}(k^\mu))\psi(k^\mu)$ requires  
\begin{equation}
e^{i{\bf \kappa}\cdot {\bf \varphi}}\eta e^{i{\bf \kappa}\cdot{\bf \varphi}}=\eta,\label{dois}
\end{equation} where $\varphi$ is the boost parameter, while invariance of the norm with respect to rotations $(\psi_{Dual}(p^\mu))\psi(p^\mu)=(\psi_{Dual}(p'^\mu))\psi(p'^\mu)$ leads to 
\begin{equation}
e^{-i{\bf \xi}\cdot {\bf \theta}}\eta e^{i{\bf \xi}\cdot{\bf \theta}}=\eta,\label{tres}
\end{equation} where $\theta$ is the rotation parameter. Eq. (\ref{dois}) leads to $\{{\bf \kappa},\eta\}=0$ whist Eq. (\ref{tres}) is satisfied when $[{\xi},\eta]=0$. These constrains allow to write $\eta$ as 
\begin{eqnarray}
\begin{bmatrix}
\mathbb{O}_{2\times 2}&\eta_1(\mathbb{I}_{2\times 2})\\
\eta_2(\mathbb{I}_{2\times 2})&\mathbb{O}_{2\times 2}
\end{bmatrix},
\end{eqnarray} where $\eta_1$ and $\eta_2$ are real parameters. Now, if $\eta_1$ and $\eta_2$ are different, then it means that we are treating differently both parts of the representation space. There is no physical reason to do so, as it would engender a parity break in the spinorial relativistic formulation. Therefore we shall consider $\eta_1=\eta_2$ and the norm is, then, fixed up to an irrelevant constant which can be absorbed into a suitable normalization. Hence $\eta=\gamma_0$ and certainly $\psi_{Dual}$ is better denoted by $\bar{\psi}$, the usual dual.

Now we are able to set down the bilinear covariants. Obviously, $\sigma=\bar{\psi}\psi$ is a scalar by construction. Moreover, denoting $j^\mu=\bar{\psi}\gamma^\mu\psi$, we have 
\begin{eqnarray}
j'^\mu=\bar{\psi}'\gamma^\mu\psi'=\bar{\psi}S^{-1}\gamma^\mu S\psi=\Lambda^\mu_{\;\;\nu}\bar{\psi}\gamma^\nu\psi,\label{vector}
\end{eqnarray} by means of Eq. (\ref{1}), showing a vectorial quantity. All the possible bilinear covariants are obtained by the same reasoning. Let us report two peculiarities: first the pseudo-quantities, i. e., the pseudo scalar $\omega=-\bar{\psi}\gamma^0\gamma^1\gamma^2\gamma^3
\psi\equiv-\bar{\psi}\gamma^{0123}\psi$ and the pseudo-vector $K^\mu=i\bar{\psi}\gamma^{0123}\gamma^{\mu}\psi$, transform as a scalar and a vector, respectively, but the transformation is always accompanied by the determinant of the Lorentz transformation, which is equal to one for orthocronoum proper transformations but is sensitive to discrete Lorentz transformations. Secondly, the last bilinear to be considered given by $S^{\mu\nu}=i\bar{\psi}\gamma^\mu\gamma^\nu\psi$ transform as a tensor but the physical bilinear associated to it is obtained by means of its contraction to the algebraic bivector $\gamma_\mu \wedge \gamma_\nu$. This last quantity represents the exterior product of base elements (the gamma's) for which the distributive rule holds but commutativity is non-longer assumed. Then, the anti-symmetrization giving rise to the usual physical coupling: 
\begin{eqnarray}
S^{\mu\nu}\gamma_\mu \wedge \gamma_\nu\!\!=\!\!\Bigg(\frac{i}{2}\bar{\psi}[\gamma^\mu,\gamma^\nu]\psi+\frac{i}{2}\{\gamma^{\mu},\gamma^{\nu}\}\!\!\Bigg)\gamma_\mu \wedge \gamma_\nu=\frac{i}{2}\bar{\psi}[\gamma^\mu,\gamma^\nu]\psi\equiv\sigma^{\mu\nu}\gamma_\mu \wedge \gamma_{\nu}  
\end{eqnarray} and, in this sense, the algebra dictates the possible couplings.

The basis of the underlying algebra, in four dimensions and mostly positive metric signature, is given by the set \cite{cliff} 
\begin{equation}
\{\mathbb{I}, \gamma_\mu, \gamma_\mu \wedge \gamma_\nu, \gamma_{0123}, \gamma_{0123}\gamma_\mu\}.\label{set}
\end{equation} Notice that, obviously not by chance, the number of base elements equals the number of bilinear covariants. Here is why: on the one hand, Clifford algebras are obtained from a suitable quotient of the exterior algebra \cite{rol}, depending in this way on the vectorial space dimension; on the other hand, the vectorial space in question is the spacetime itself. Thus, the algebra is determining the possible amount of couplings.  Just a parenthetical remark, as the covariant bilinears are associated to physical observables, it is important to ensure that all of them are real quantities. In fact it can be always guaranteed with the aid of precise deformations of the base (\ref{set}) \cite{craw1}. 

Provided with the base and all the bilinear, we are in position to appreciate an important element in the Lounesto classification. It is the so-called Fierz aggregate. Here we shall give less importance to this quantity then it deserves, only pinpointing its main aspects to our purposes, but the interested reader may found a complete account on that in Ref. \refcite{lou}. The Fierz aggregate, $Z$, is given by the sum of the base contracted bilinear covariants: 
\begin{eqnarray}
Z=\sigma + j^\mu \gamma_\mu + iS^{\mu\nu}\gamma_\mu \wedge \gamma_\nu + iK^\mu \gamma_{0123} \gamma_\mu + \omega\gamma_{0123}.\label{ze}
\end{eqnarray} In some sense (\ref{ze}) contain all the relevant information about the physical aspects of a spinor based theory. Its importance to our program rest on the fact that a given spinor, $\psi$, which gave rise to the covariant bilinears (and therefore to $Z$ itself) can be (re)obtained from $Z$. This is a strong statement, whose proof lies along a constructive method \cite{lou}. Taking the result, we are able to write
\begin{eqnarray}
\psi\sim Z e^{-i\varphi}\eta, \label{psi}
\end{eqnarray} where $\eta$ is an arbitrary constant spinor and $\varphi$ a phase.  We shall not be concerned to the proof here, but we would like to give a circumstantial argument in favor of (\ref{psi}).
  
Remember that a given spinor is endowed with four complex entries, and therefore there are eight real {\it degrees of freedom}. By inspecting $Z$ one see that it has 1 ($\sigma$) $+$ 4 ($j^\mu$) $+$ 6 (anti-symmetric $S^{\mu\nu}$) $+$ 4 ($K^\mu$) $+$ 1 ($\omega$) degrees of freedom. Additionally, the phase also contributes to one more degree of freedom. Hence, we are left with 17 degrees of freedom in the right-hand side of (\ref{psi}) where only eight were expected. In other words, if Eq. (\ref{psi}) is valid, then the bilinear covariants must be constrained. And in fact they are: the bilinear quantities respect certain quadratic relations, the FPK identities \cite{fierz,pauli,kof}. These identities read 
\begin{eqnarray}
j_\mu j^\mu=\sigma^2+\omega^2, \label{pri} \\ K_\mu K^\mu=-j_\mu j^\mu, \label{seg}\\ j_\mu K^\mu=0, \label{ter}\\ \sigma_{\mu\nu}=\frac{1}{\sigma^2+\omega^2}\{\sigma \epsilon_{\mu\nu\alpha\beta}j^\alpha K^\beta-\omega (j_\mu K_\nu-K_\nu j_\mu)\}, \label{qua} 
\end{eqnarray} where $\epsilon_{\mu\nu\alpha\beta}$ is the Levi-Civita completely anti-symmetric symbol with convention adopted such that $\epsilon_{0123}=-1$. Notice that the set of Eqs. (\ref{pri})-(\ref{qua}) constraints nine degrees of freedom and, then, the Fierz aggregate is left with only seven ($16-9$) degrees of freedom. Thus, taking into account the phase we have a perfect balance in Eq. (\ref{psi}). It is important to remark that the aforementioned Fierz-Pauli-Kofink equations are only valid in the so-called regular case, i. e. the case in which the spinor gives rise to simultaneously non-vanishing $\sigma$ and $\omega$ (see Eq. (\ref{qua})). Nevertheless, it is possible to work out the previous set of equation and replace it for a more general set valid in every possible case:  
\begin{eqnarray}
\sigma_{\mu\nu}K^\nu=\omega j_{\mu}, \hspace{.3cm} \frac{1}{2}\epsilon_{\mu\nu\alpha\beta}\sigma^{\alpha\beta}K^{\nu}=\sigma j_{\mu},\label{sing1}\\
\sigma_{\mu\nu}j^\nu=\omega K_{\mu}, \hspace{.3cm} \frac{1}{2}\epsilon_{\mu\nu\alpha\beta}\sigma^{\alpha\beta}j^{\nu}=\sigma K_{\mu},\label{sing2}\\
\frac{1}{2}\sigma\epsilon_{\mu\nu\alpha\beta}\sigma^{\alpha\beta}+\omega\sigma_{\mu\nu}=(K_\mu j_\nu-j_\mu K_\nu), \hspace{.3cm} \sigma\sigma_{\mu\nu}=\epsilon_{\mu\nu\alpha\beta}j^\alpha K^\beta,\label{sing3}\\
\sigma_{\mu\nu}\sigma^{\mu\nu}=2(\sigma^2-\omega^2),\label{sing4}\\
\frac{1}{2}\epsilon_{\mu\nu\alpha\beta}\sigma^{\alpha\beta}\sigma^{\mu\nu}=-4\sigma\omega.\label{sing5}
\end{eqnarray} It is straightforward to see that this last set of equations constraints the Fierz aggregate degrees of freedom in the very same way the previous identities does. 

The idea behind the Lounesto classification scheme is the following: since Eq. (\ref{psi}) is valid, then every different combination of the observables, provided the FPK are respected (which is indeed a condition for Eq. (\ref{psi})), may lead to a different spinor class. But different values for a given bilinear, alone, is not enough to categorize different spinor fields, except when one of the values is zero. In fact, a null bilinear deserves a different classification, since it eliminates the physical observable associated to the vanishing bilinear. This reasoning lead us to the last aspect of the Lounesto spinor classification: the bilinear $j^\mu$ can never be null. The reason is quite clear. Bearing in mind the classification restricted to the usual fermionic fields, then the Dirac equation must dictate the dynamics to be obeyed by the field. Therefore, $j^\mu$ is nothing but the conserved current. The vanishing of this bilinear would, essentially, implicate the non-existence of the fermionc relativistic particle one is trying to describe with the spinor at hand. Hence a non-vanishing $j^\mu$ is the last imposition to the classification procedure.  

Let us first concentrate in the case in which $\sigma$ and $\omega$ are not simultaneously null, i. e. the regular case. In this case we are well supported by Eqs. (\ref{pri})-(\ref{qua}) and Eq. (\ref{pri}) ensures $j^\mu\neq 0$. From Eqs. (\ref{seg}) and (\ref{qua}) we have no option then $K^\mu$ and $\sigma^{\mu\nu}$ different from zero\footnote{Just for further  convenience, remember that a given non-scalar quantity $\chi^2\neq 0$ implies $\chi \neq 0$, but obviously $\chi^2=0$ does not mean $\chi=0$ (necessarily) in a pseudo-Euclidean space.}. Hence we are left with only three different possibilities:

(1) $\sigma\neq 0$ and $\omega\neq 0$;

(2) $\sigma\neq 0$ and $\omega=0$;

(3) $\sigma=0$ and $\omega\neq 0$. 

For all these cases it is possible to have information about the distribution $(\sigma^{\mu\nu})$ and direction $(K^\mu)$ of the (electron) intrinsic angular momentum. Having exhausted the possibilities concerning regular spinors, we now move to the singular cases, the cases for which $\sigma=0=\omega$. Now we can only trust in Eqs. (\ref{sing1})-(\ref{sing5}). Notice that once we implement $\sigma=0=\omega$ and the fact that $j^\mu$ cannot be zero, then we are again left with three, and only three, cases. This can be seen from the following reasoning. First, consider $K^\mu \neq 0$. Within this premise we note that there is no constraint to be imposed to $\sigma^{\mu\nu}$ coming from our set of equations. It is to be emphasized that Eqs. (\ref{sing1})-(\ref{sing5}) do not perform the best set to study the singular case \cite{lou}. Nevertheless, we shall keep our presentation as it is for pedagogical reasons. It is straightforward to see that the cases below are in order 
 
(4) $K^\mu\neq 0$ and $\sigma^{\mu\nu}\neq 0$; 

(6) $K^\mu \neq 0$ and $\sigma^{\mu\nu}=0$.

The unusual enumeration adopted is just to keep track to the Lounesto original classification \cite{lou}. For spinors belonging to class (4), it is fairly simple to see that $\sigma_{\mu\nu}j^\nu=0=\sigma_{\mu\nu}K^\nu$. Therefore $\sigma_{\mu\nu}(j^\nu-K^\nu)=0$ and as $\sigma_{\mu\nu}\neq 0$ for $\mu\neq\nu$ then $j^\mu=hK^\mu$ for a constant $h$, giving rise to the helicity concept in mathematics \cite{boe}. In this case Eq. (\ref{sing3}) is automatically satisfied. Before to evince a parallel between the mathematical denomination applied to some classes and the bilinear covariants, let us explore the last possibility according to this classification. Setting $K^\mu=0$, Eqs. (\ref{sing2}) and (\ref{sing3}) are readily satisfied, and again we have no restriction to impose on $\sigma_{\mu\nu}$. Hence, one possibility is giving by 

(5) $K^\mu=0$ and $\sigma^{\mu\nu}\neq 0$.     

One may wonder about a possible class giving by vanishing $K^\mu$ and $\sigma^{\mu\nu}$ at the same time. It seems a logical possibility indeed. However, notice that in this case the Fierz aggregate would be entirely determined by a single quantity, the conserved current, and from Eq. (\ref{psi}) we have $\psi^\dagger\psi=(Z\eta)^\dagger(Z\eta)=0$ for this singular tentative case. Thus, $\eta^\dagger \gamma_\mu^\dagger j^\mu j^\alpha\gamma_\alpha \eta=0$ and since $\eta$ must be non-null one would be forced to conclude that $j^\mu=0$ and the spinor itself is vanishing, a clear contradiction. Hence, the Fierz aggregate cannot be only composed by a single bilinear only, and we are left with six and only six classes.  

The three first classes are usually connected to the relativistic physics for the electron. For this reason Lounesto called the elements belonging to these classes by Dirac spinors. As we mentioned, a bivector is an element of the Clifford algebra basis. As an exterior product, the bivector and the vectorial product have a quite similar interpretation. In this vein it is usual to associate the bivector to an oriented plane defined by its individual vectors. Mathematicians like to call it a flag, while a single vector is connoted by a pole (for a precise account on the formal aspects of this designations, see Ref. \refcite{fp}). Therefore, as  class (4) elements have $j^\mu\neq 0$ (a pole), $K^\mu\neq 0$ (another pole) and $\sigma^{\mu\nu}\neq 0$ (a flag), they are called Flag-dipole spinors. 
According to the previous designation, elements of class (5) are called Flag-pole spinors. For instance, Majonara spinors can be recast into this class. Spinors of class (6) are the Weyl spinors, the very eigenspinors of the chirality operator.  

We would like to finalize this section making a remark about the classification of Elko \cite{dva2} and other spinors \cite{dva} whose formulation needs a different dual. A previous classification for Elko allocate\cite{rol2} these spinors as class (5). Strictly speaking, the classification according to Lounesto needs a dual {\it a la} Dirac and a new classification scheme must be taken into account for different duals. However, once one are aware of this fact, the study of these different dual spinors in the above context may also be informative in the sense that one would be interested in algebraically investigate the position, in a manner of speaking, of these spinors according to the Lounesto scheme.   

\section{Shortcuts for further results}\label{shortcuts}

The content of the previous section represents the basis upon 
the Lounesto classification was build up. However, the subject has been investigated and developed since then, being very active in the last few years. The aim of this section is to organize the main recent results as a guide for non specialists interested in the field. Following the guideline of the central role played by the bilinear covariants, we have chosen 11 papers that represent the advances in \textit{applying}\cite{rol2,daRocha:2007pz,ijmpa,nos,daRocha:2016bil}, \textit{understanding}\cite{Cavalcanti:2014wia,daSilva:2012wp} and \textit{extending}\cite{Ablamowicz:2014rpa,Villalobos:2015xca,Bonora:2014dfa,deBrito:2016qzl,Bonora:2015ppa} the Lounesto seminal result. Such stages naturally appeared in a roughly  chronological order.
It is worth  emphasizing that, besides Elko spinors do not exactly fit in any of the Lounesto classes, it has been motivating a considerably amount of research in the field of spinor fields classification.  It was particularly true for the initial investigations, as we shall discuss below. A proper definition of bilinear covariants regarding the Elko dual particularities and obeying the FPK identities was found in Ref. \refcite{HoffdaSilva:2016ffx}. 

\subsection{Applying}

This preliminary stage of the Lounesto classification development is characterized by classifying spinors appearing outside the domain of Dirac theory. In doing so, as mentioned before, Ref. \refcite{rol2} classified Elko spinor field as flagpole. Regardless of the Elko alternative dual, the result was important for bringing the Lounesto classification, for the first time, to an audience wider then the Clifford algebra community. The authors also showed that every flagpole spinor field is composed by two 2-spinors with opposite helicities. At this time the class (4) were the only one whose elements had not explicitly appeared in any physical theory. Ref. \refcite{daRocha:2007pz}, on the other hand, investigate the possibility of constructing maps from Dirac to Elko spinors. 
Such maps were proposed as an attempt to relate Elko and Dirac dynamics and, preliminarily, investigate a way for extending the standard model in order to encompass Elko spinors. The results were found, at first, by considering a general map $M$ such that $M\psi=\lambda$. As usual, $\psi$ represents a Dirac spinor and $\lambda$ denotes an Elko spinor. An invertible map was found, allowing one to write down an Elko spinor in terms of components of a Dirac spinor. The next step was find the constrains that components of spinors of the first three classes should obey in order to keep the vanishing condition of bilinear covariants of the class (5). It established general conditions for a Dirac spinor be mapped into Elko spinors. This mapping was further applied to evince a symmetry between Elko and Dirac actions\cite{ijmpa}.

A representative of the class (4) explicitly appeared as a physical solution for the first time in Ref. \refcite{nos}. The paper analysed solutions of the Dirac equation in the Einstein-Sciama-Kibble theory of gravity, which is a class of $f(R)$ theories including torsion. It was shown that, in such environment, the solutions of the Dirac equations are not restricted to the classes (1), (2) and (3). Two sets of field equations were derived in the theory, the first one analogous to the Einstein equation in $f(R)$ theories and another one coupling the torsion tensor to the spin density tensor. Assuming an axially symmetric Bianchi-I type for the background metric, given by $ds^2=dt^2-a(t)dx^2-b(t)dy^2-c(t)dz^2$, the corresponding Dirac equation was shown to have solutions of the form 
\begin{align}\label{eks}
\psi=\frac{1}{\sqrt{2\tau}}\begin{pmatrix}
\sqrt{K-C}\cos \zeta_1e^{i\theta_1}\\
0\\
0\\
\sqrt{K+C}\sin \zeta_2e^{i\theta_2}
\end{pmatrix} \quad\mbox{ and }\quad \psi=\frac{1}{\sqrt{2\tau}}\begin{pmatrix}
0\\
\sqrt{K+C}\cos \zeta_1e^{i\vartheta_1}\\
\sqrt{K-C}\sin \zeta_2e^{i\vartheta_2}\\
0
\end{pmatrix},
\end{align}
where $\tau=a(t)b(t)c(t)$, $\zeta_1, \zeta_2, \theta_1,\theta_2,\vartheta_1$ and $\vartheta_2$ are time dependent parameters and  $C,K$ are constants. As we shall discuss in the next section, the spinors in Eq. \eqref{eks} have exactly the form of  class (4) spinors with two non null components. A remarkable result concerning the Lounesto classification here is that, under the right conditions,  generalizations of the Dirac equation do allow singular spinors as physical solution.

 Following the fruitful route connecting spinor fields and gravity, Refs. \refcite{Mei:2010wm} and \refcite{Mei:2011gv} introduced properties of a fluid description dual to rotating black hole solutions. In such approach the vorticity is related to the exterior derivative of a fluid flow and
the Riemann curvature tensor, being coupled through the spin density. This coupling is described by
\begin{align}
d(\rho u)=i\bar{\psi}\gamma^{\mu}\gamma^{\nu}\psi R_{\mu\nu\lambda\sigma}dx^\lambda \wedge dx^\sigma\equiv S^{\mu\nu} R_{\mu\nu\lambda\sigma}dx^\lambda \wedge dx^\sigma,
\end{align}
where $\rho$ is the fluid density and $u$ the four velocity ($u^\nu u_\nu=-1$). The  possibility of the above fluid description being generated be singular spinors of the classes (4) and (5) was analysed in Ref. \refcite{daRocha:2016bil}. It was shown that, considering the Kerr solution, the fluid description can be generated by singular spinors placed near the black hole horizon. Is was the second example, up to our knowledge, of a physical model having  class (4) spinors as solution.

\subsection{Understanding}

At this time the reader might be noticed that the classification of spinors, according to the Lounesto proposal, involves a large amount matrix computations that rather than technically difficult are quite time demanding. One of the aims of Ref. \refcite{Cavalcanti:2014wia} was turn the classification process as straightforward as possible, characterizing all the classes by converting the general constraints on the bilinear covariants into general constraints on components of spinors of each class. A table with the most general form of spinors of all singular classes was build up based on the number of non null components. Thus, given an arbitrary spinor filed $\psi=(\phi_1,\phi_2,\phi_3,\phi_4)^\intercal$, where $\phi_i:\mathbb{R}^{1,3}\to \mathbb{C}$, its classification is achieved by doing calculations as simple as the norm of a couple of complex numbers. Besides the considerable simplification on the classification process, these results could also be useful on investigations of the singular spinors dynamics. The characterization results, assuming the Weyl representation for the $\gamma$ matrices, are reproduced in the Table \ref{tab1} below:
\begin{table}[h]
\centering
\tbl{Spinors characterization table. *NNNC means number of non null components.}{
\begin{tabular}{ccccc}
\toprule
Classes & Type-(4) & Type-(5) & Type-(6) & Regular \\ 
\colrule
*NNNC &  &  &  &  \\
1 & -- & -- & Arbitrary & -- \\ \\
2 & $\begin{pmatrix}
\phi_1\\
0\\
0\\
\phi_4
\end{pmatrix},\;\begin{pmatrix}
0\\
\phi_2\\
\phi_3\\
0
\end{pmatrix}$&$\begin{pmatrix}
\phi_1\\
0\\
0\\
\phi_4
\end{pmatrix},\;\begin{pmatrix}
0\\
\phi_2\\
\phi_3\\
0
\end{pmatrix}$&$\begin{pmatrix}
\phi_1\\
\phi_2\\
0\\
0
\end{pmatrix},\;\begin{pmatrix}
0\\
0\\
\phi_3\\
\phi_4
\end{pmatrix}$ & $\begin{pmatrix}
\phi_1\\
0\\
\phi_3\\
0
\end{pmatrix},\;\begin{pmatrix}
0\\
\phi_2\\
0\\
\phi_4
\end{pmatrix}$\\ 
&$\begin{array}{l}
 ||\phi_1||^2 \neq ||\phi_4||^2\\
  ||\phi_4||^2 \neq ||\phi_3||^2
\end{array}$&$\begin{array}{l}
 ||\phi_1||^2 = ||\phi_4||^2\\
  ||\phi_2||^2 = ||\phi_3||^2
\end{array}$&&\\ \\
3 & -- & -- & -- & Arbitrary \\ \\
4 & $\begin{pmatrix}
-\frac{\phi_2\phi_3\phi_4^*}{||c||^2}\\
\phi_2\\
\phi_3\\
\phi_4
\end{pmatrix}$ &
$\begin{pmatrix}
-\phi_4^*e^{i \varphi}\\
\phi_3^*e^{i \varphi}\\
\phi_3\\
\phi_4
\end{pmatrix}$ & --& $\begin{pmatrix}
\phi_1\\
\phi_2\\
\phi_3\\
\phi_4
\end{pmatrix}$ \\ 
&$||\phi_2||^2 \neq ||\phi_3||^2$&&&$\phi_1 \neq -\frac{\phi_2\phi_3\phi_4^*}{||\phi_3||^2}$\\ 
\botrule
\end{tabular}} 
 \label{tab1}
\end{table}

The possibility of finding spinor fields obeying dynamical equations different to the one proposed by Dirac brought back a fundamental question concerning the uniqueness of the spinorial structure. According to the mathematical theory of the Dirac operator, if the space has non trivial topology, in the sense of non trivial fundamental group $\pi_1$, the Dirac equation is not unique. In such cases an additional term emerge correcting the dynamical equations. Solutions of those  equations are known as exotic spinor fields and were related to the Lounesto classification in Ref. \refcite{daSilva:2012wp}. Fundamental investigations connecting exotic spinors and sources of non trivial topology are under investigations. The subject is very technical and out of the scope of the present paper, nevertheless a comprehensive discussion on the topic can be found in Refs. \refcite{daRocha:2011yr}, \refcite{Geroch:1968zm} and \refcite{Fthomas}.

\subsection{Extending} \label{ext}

In spite of a background in Clifford algebras being not required for understanding and applying the Lounesto spinors classification, it is not always true for its recent extensions.   The results described in the present subsection extend the classification in different ways, including quantum  Clifford algebras \cite{Ablamowicz:2014rpa}, relaxation of  the constrain $j^\mu \neq 0$ \cite{Villalobos:2015xca} and extensions to spinors in extra dimensional spacetimes \cite{Bonora:2014dfa,deBrito:2016qzl,Bonora:2015ppa}. Regardless the fact of the mentioned results being largely rooted on Clifford algebras, we shall keep using only the essential results and definitions of such structure.  Our aim is to avoid technical details in favor of a wider view of recent investigations concerning spinors classifications based upon bilinear covariants. The reader interested on the algebraic structure of spinors, Clifford algebras and its applications in physics will certainly find Refs. \refcite{lou,rol} very useful.

Clifford algebras are rich algebraic structures defined over quadratic linear spaces, or equivalently, over linear spaces endowed with symmetric bilinear forms, by $\{\gamma_\mu,\gamma_\nu\}=2g_{\mu\nu}\mathbb{I}$.  Here $\{\cdot,\cdot\}$ denotes the anti-commutator, $\gamma_\sigma$ are the algebra generators,  $g_{\mu\nu}$ is the  symmetric bilinear form and $\mathbb{I}$ is the algebra unity. Quantum Clifford algebras, on the other hand, are similarly defined over linear spaces endowed with non symmetric bilinear forms. It allows an even richer algebraic structure, whose applications in physics varies from quantum field theory to statistical mechanics (see Ref. \refcite{Ablamowicz:2014rpa} and references therein). The important fact for our proposes here is that spinors and bilinear covariants can be defined over quantum Clifford algebras analogously to the ones defined over standard Clifford algebras. It gives a natural generalization of the Lounesto classification based on the new bilinear covariants. In fact, this is the main result introduced in Ref. \refcite{Ablamowicz:2014rpa}, where the new quantum classes were derived and thoroughly analysed. Finding representatives of those classes is an interesting problem which remains open.

Clifford algebras are also the natural algebraic structure suitable for defining spinors. Besides the classical definition introduced in Sec. \ref{LCS}, spinors can be defined by two alternative and equivalent ways, both independent of any choice of representation and constructed merely upon the basic structure of Clifford algebras\cite{rol}. According to the so called algebraic definition, spinors are elements of minimal left ideals of the form $\mathcal{C}\ell_{1,3}f$, where $\mathcal{C}\ell_{1,3}$ denotes the Clifford algebra of the Minkowski spacetime and $f$ is a primitive idempotent. The operatorial spinors, on the other hand, are defined as elements of the even sub algebra of $\mathcal{C}\ell_{1,3}$, denoted by $\mathcal{C}\ell_{1,3}^+$. The Takahashi theorem, the FPK identities and their relationship with bilinear covariants were investigated in the context of algebraic and operatorial spinors in Ref.  \refcite{Villalobos:2015xca}. The motivation was the fact that, even though the bilinear covariant $j^\mu$ being interpreted as the probability current density within the Dirac theory, such interpretation is not valid when the dynamics is not described by the Dirac equation, which is the case for singular spinors. Thus, in principle, $j^\mu \neq 0$ is not required in the general case. In fact, it was shown that the above spinors definition are perfectly consistent without imposing $j^\mu \neq 0$ \cite{Villalobos:2015xca}. Thereafter the Lounesto classification is straightforwardly extended by adding the constrain $j^\mu = 0$, revealing three additional classes, namely:
\begin{enumerate}
\item[(4a)]  $\sigma = 0 = \omega,\; j^\mu = 0,\; K^\mu\neq 0$ and $S^{\mu\nu}\neq 0$

\item[(5a)]  $\sigma = 0 = \omega,\; j^\mu = 0,\; K^\mu= 0$ and $S^{\mu\nu}\neq 0$
\item[(6a)]  $\sigma = 0 = \omega,\; j^\mu = 0,\; K^\mu\neq 0$ and $S^{\mu\nu}= 0$
\end{enumerate}
%
%

The last type of extension of the Lounesto classification we are going to introduce is the one related to extra dimensional spacetimes. Whether it comes from fundamental string/M-theory or braneworld models, the possibility of extra dimensional spacetimes has undeniably generated a large amount of investigation in the last two decades. It makes a careful investigation of possible types of spinors existing in those spacetimes more than deserved. Despite the intricate geometric and algebraic structure of higher dimensional spinors being a true challenge, after finding the bilinear covariants the procedure for constructing the classes is, in principle, quite similar to the one we have been discussing. The Fierz identities generalized to spacetimes with an arbitrary number of dimensions were already known \cite{Lazaroiu:2013kja}, thus Ref. \refcite{Bonora:2014dfa} established the Bilinear covariants for such higher dimensional spaces. In addition, motivated by the AdS$_4\times S^7$ model of supergravity, a detailed analysis of the spinors classification on the sphere $S^7$ was performed. They also found that the Fierz identities forbid the existence of more than one spinor field class for the real Majorana spinors on Riemannian 7-manifolds. For the general complex case the number of classes increases to three. Those investigations were generalized to Lorenzian manifolds in Ref.  \refcite{Bonora:2015ppa}, were the obstruction for new classes found in the Riemannian case was shown to not be always present for Lorenzian signatures. The case of spinors in 5 dimensional braneworld models was studied in Ref. \refcite{deBrito:2016qzl}, touching the problem of fermions localization and the new classes therein. We finish by arguing that, taking Elko spinors as example,  finding representatives of all those new classes is an open problem whose implications are potentially connected to fundamental questions of the 21st. century physics.


\section{Final remarks}

In the present paper we have revisited a powerful and comprehensive program of classification of spinors initiated by Pertti Lounesto, the so called Lounesto spinors classification. The fundamental algebraic structure behind the classification, as the bilinear covariants, the Takahashi theorem, the FPK identities and the Fierz aggregate were introduced in a clear way. After a careful discussion on the fundamental details of the original scheme of classification,  the further results were revised. It includes applications, simplifications and generalizations of the original spinors classification. The new classes of spinors coming from those generalizations are still lacking physical realisation of their representatives, proposing interesting problems for future researches. It is worth mentioning that those results are very recent (2015/2016) and under development. In addition, it took twelve years for finding the first type (4) representative\cite{nos} after the Lounesto original results were compiled in Ref. \refcite{lou}.

The general method introduced by Lounesto has lead to several developments in the understanding of spinors and its use in physics. Moreover, it opened possible lines of investigation, giving additional building blocks for new (including quantum) fields. In a four percent known universe, it seems a program to remember! 

\section*{Acknowledgments}

The authors thanks the always fruitful conversations with Professor Roldao da Rocha. JMHS thanks CNPq, grants number (304629/2015-4; 445385/2014-6), for financial support. RCT thanks the UNESP-Guaratinguet\'a Post-Graduation program and CAPES.

\end{document}